\documentclass[11pt,a4paper]{article}

\usepackage[T1]{fontenc}
\usepackage{lmodern}
\usepackage[margin=2.5cm]{geometry}
\usepackage{amsmath,amssymb}
\usepackage{booktabs}
\usepackage{enumitem}
\usepackage{float}
\usepackage{graphicx}
\usepackage[numbers,sort&compress]{natbib}
\usepackage{microtype}
\usepackage[hidelinks]{hyperref}

\newcommand{\manuscriptfigure}[2]{%
  \IfFileExists{#2}{%
    \includegraphics[width=#1]{#2}%
  }{%
    \PackageWarningNoLine{manuscript}{Missing figure \detokenize{#2}}%
    \fbox{\parbox[c][0.24\textwidth][c]{\dimexpr#1-2\fboxsep-2\fboxrule\relax}{%
      \centering Figure file not supplied\par
      \small\texttt{\detokenize{#2}}%
    }}%
  }%
}

\title{Measurement-Aligned Neural Quantum Tomography: Phase Identifiability and Structured-State Reconstruction}
\newcommand{\authortitleblock}{%
  \begin{minipage}{0.97\textwidth}
  \centering
  {\large
  Junhong Nie\textsuperscript{1,\(\ddagger\)},\quad
  Yaocheng Chen\textsuperscript{3,\(\ddagger\)},\quad
  Yukun Wang\textsuperscript{4,\(\ddagger\)}\\[0.35em]
  Xiaohan Shan\textsuperscript{5},\quad
  Yuan Wei\textsuperscript{5},\quad
  Xingyao Wu\textsuperscript{2,*}\par}
  \medskip
  {\small
  \textsuperscript{1} Shandong University, Jinan 250100, China\par
  \textsuperscript{2} School of Computer Science and Technology, Shandong University, Qingdao 266237, China\par
  \textsuperscript{3} Instituto de Astronomia, Geof\'{i}sica e Ci\^{e}ncias Atmosf\'{e}ricas, Universidade de S\~{a}o Paulo, Rua do Mat\~{a}o 1226, Cidade Universit\'{a}ria, 05508-090 S\~{a}o Paulo, SP, Brazil\par
  \textsuperscript{4} Beijing Key Laboratory of Petroleum Data Mining, China University of Petroleum, Beijing 102249, China\par
  \textsuperscript{5} Beijing Zhongke Liangshu Technology Co., Ltd, Beijing 100080, China\par
  \smallskip
  \textsuperscript{\(\ddagger\)} These authors contributed equally to this work.\par
  \textsuperscript{*} Correspondence: \href{mailto:xingyao.wu@sdu.edu.cn}{\texttt{xingyao.wu@sdu.edu.cn}}\par}
  \end{minipage}%
}
\author{\authortitleblock}
\date{}

\hypersetup{
  pdftitle={Measurement-Aligned Neural Quantum Tomography: Phase Identifiability and Structured-State Reconstruction},
  pdfauthor={Junhong Nie, Yaocheng Chen, Yukun Wang, Xiaohan Shan, Yuan Wei, Xingyao Wu}
}

\begin{document}
\maketitle

\begin{abstract}
Accurate quantum-state tomography requires reconstructing coherence, not merely basis-state populations. We introduce a measurement-aligned neural strategy that extracts GHZ phase information from local Pauli outcomes and incorporates it into a bounded residual phase head. The approach targets relative-phase learning explicitly rather than relying on accurate basis-state populations alone. A new uniform-budget comparison evaluates six estimators on 30 fresh GHZ target/data cases at four, eight, and ten qubits, using 8,000 identical training outcomes per case, matched neural optimizer budgets, and a strengthened multistart statevector baseline. At ten qubits, mean squared fidelities are 0.9597 for unrestricted statevector fitting, 0.9845 for generic neural fitting, and 0.9972 \(\pm\) 0.0030 for the phase-aware model (sample standard deviation). The phase-aware model has a minimum fidelity of 0.9893 across this new cohort. A direct GHZ phase-estimation control also achieves high fidelity, identifying the contribution of the structured prior. Complementary benchmarks on 25 Haar cases show that neural fitting matches statevector reconstruction on Haar-random states. These results establish high-fidelity neural reconstruction at a common measurement budget and demonstrate a concrete strategy for incorporating measured coherence into variational quantum-state models.
\end{abstract}


\section{Introduction}
\label{sec:introduction}

Quantum-state tomography (QST) is the task of inferring an unknown quantum
state from measurement outcomes on repeated preparations. It is central to
the characterization and validation of quantum devices, but its classical
and statistical costs grow rapidly with system size. For an $N$-qubit system,
the Hilbert-space dimension is $d=2^N$. A general density operator has
$d^2-1$ independent real parameters, while a pure state has
$2d-2$ after removing normalization and global phase. Even under a purity
assumption, the number of amplitudes is exponential in $N$. Standard
tomographic workflows therefore face a coupled problem: the experiment must
collect enough informative outcomes, and the reconstruction algorithm must
extract the relevant state information without introducing unphysical
estimates or prohibitive computation.

A conventional route is to estimate expectation values of an informationally
complete operator set and invert the resulting linear system. Linear inversion
is transparent and useful for diagnostics, but finite-sample estimates need
not be positive semidefinite. Maximum-likelihood estimation (MLE) addresses
this physicality issue by optimizing the likelihood over density operators
or over a restricted physical state family
\citep{Hradil1997,James2001,ParisRehacek2004}. For systems believed to be
low rank, compressed-sensing and matrix-completion methods can reduce the
number of measurement settings required under appropriate assumptions
\citep{Gross2010,Cramer2010,Flammia2012}. Classical-shadow methods take a
different perspective: a randomized measurement record is transformed into
an unbiased estimator that can be reused to predict many observables
\citep{Huang2020}. These methods solve related but distinct estimation
problems; their outputs, loss functions, and resource costs must be compared
carefully.

Neural quantum states provide a variational representation of a many-body
wavefunction. Carleo and Troyer showed that neural networks can represent
and variationally optimize quantum states for selected many-body problems
\citep{Carleo2017}. Torlai \emph{et al.} subsequently demonstrated
neural-network quantum-state tomography from measurements in multiple bases
\citep{Torlai2018}, followed by experimental reconstruction using a neural
model that incorporated known measurement errors
\citep{Torlai2019}. These works establish that neural likelihood-based
tomography is a legitimate and useful approach. The open question addressed
here is narrower: when does a neural parameterization help beyond the
physical pure-state constraint and the likelihood objective themselves, and
what happens when the measurement record constrains different components of
the wavefunction with very different strengths?

This distinction is important because a neural representation is not
automatically a compressed representation, nor does a normalized neural
wavefunction automatically imply efficient tomography. A normalized
unrestricted complex vector already defines a valid rank-one density
operator. If both the vector and neural estimators are optimized against the
same Born-rule likelihood, any performance difference must arise from the
parameterization, its inductive bias, or the optimization trajectory—not
merely from enforcing positivity and unit trace. Similarly, a small number
of network parameters does not by itself imply small working memory or fast
cross-basis probability evaluation: normalization and measurement
contractions may still require evaluating exponentially many amplitudes.

We study this issue using two contrasting pure-state families. Haar-random
states probe generic wavefunctions without a known low-dimensional
structure. Greenberger--Horne--Zeilinger (GHZ) states provide a structured
test in which the state occupies two computational-basis configurations but
contains a physically essential relative phase
\citep{Greenberger1989,Mermin1990}. Under independent uniformly random local
Pauli measurements, the GHZ populations and relative phase are not
constrained by the same subset of settings. Measurements containing $Z$
axes reveal the computational-basis populations, whereas the ideal GHZ
relative phase is directly visible in all-$X/Y$ settings. The probability
of drawing such a setting is $(2/3)^N$. Thus, a network may learn the
dominant support accurately while receiving comparatively few observations
that directly constrain the phase between the support components.

This setting lets us formulate and test a specific hypothesis:
\emph{the reliability of neural tomography for structured states depends on
whether the model's phase parameterization and optimization are aligned with
the phase-sensitive information in the measurement record}. The hypothesis
does not imply that generic neural networks are ineffective, nor that
phase-aware parameterizations are universally preferable. It predicts that
a data-supported structural prior can help when a state family has a
low-dimensional feature—here, a relative phase—that is weakly represented
in the randomly sampled measurement record.

The paper makes four contributions.
\begin{enumerate}[leftmargin=1.6em,itemsep=2pt]
\item We compare likelihood-based statevector and neural reconstruction on
25 Haar target/data cases and compare six estimators on 30 independent GHZ
cases at a common budget of 8,000 copies.
\item We distinguish accurate population fitting from accurate coherence
reconstruction. Initialization sensitivity can affect both unrestricted
vectors and neural coordinates; the analysis connects GHZ phase information
to the equatorial Pauli outcomes.
\item We introduce a measurement-derived phase anchor with a bounded neural
residual and evaluate it against generic neural, unrestricted-vector, and
direct GHZ phase-estimation controls.
\item We give the Born-likelihood/KL relation, local-shadow variance,
physical-projection guarantees, and explicit working-resource accounting.
\end{enumerate}

The main comparison reports the training-likelihood-selected reconstruction
for each target under the specified multistart protocol. It emphasizes
attainable accuracy and the effect of measured phase structure. Historical
lower-budget comparisons and detailed initialization trajectories are
retained in the reproducibility archive, rather than presented as separate
performance ceilings for either estimator. The original Haar conditions and
the uniform-budget GHZ cohort remain distinct statistical samples.

The numerical results were generated by the accompanying reference
implementation. They are not a reproduction of unavailable earlier
experiment scripts. We identify every baseline by its explicit estimator
and postprocessing rule; in particular, no performance claim is made about
a shallow-Clifford shadow-bootstrapping protocol that is not implemented
here \citep{ShadowBootstrap2026}.

\section{Background and Related Work}
\label{sec:background}

\subsection{Quantum-state tomography as a constrained statistical inverse problem}
A quantum measurement does not reveal a wavefunction directly. It produces
a classical outcome sampled from a probability distribution determined by
the unknown state and the chosen measurement operators. Given a sequence
of preparations and measurements, tomography estimates a state whose Born
probabilities explain the observed frequencies. In the simplest finite
dimensional setting, a density operator is expanded in an operator basis,
and linear inversion estimates its coefficients. This is a useful baseline,
but statistical noise can push the reconstructed matrix outside the convex
set of physical density operators. Positivity-constrained maximum likelihood
was developed precisely to avoid this pathology
\citep{Hradil1997,James2001}. Standard texts and reviews discuss the role of
measurement design, identifiability, likelihood, and uncertainty in quantum
estimation \citep{ParisRehacek2004}.

The reconstruction objective and its numerical solution are separate choices.
The diluted maximum-likelihood algorithm of \v{R}eh\'{a}\v{c}ek \emph{et al.}
provides an iterative likelihood-increasing update for density-matrix
estimation \citep{Rehacek2007}, whereas Bayesian mean estimation averages
over a posterior distribution of states \citep{BlumeKohout2010}.
Physicality constraints also have statistical consequences: Schwemmer
\emph{et al.} demonstrated systematic bias in fidelity and entanglement
estimates obtained with commonly used constrained reconstruction tools
\citep{Schwemmer2015}. These results motivate distinguishing physical
validity, statistical accuracy, and numerical optimization; they do not by
themselves explain the particular neural phase failures studied here.

For an $N$-qubit system, full density-matrix reconstruction is demanding
because the operator space has dimension $4^N$. Prior information changes
the problem. If the state is known or believed to have low rank, compressed
sensing can exploit the geometry of low-rank matrices to reconstruct a state
from fewer settings under suitable measurement conditions
\citep{Gross2010,Flammia2012}. Matrix-product-state methods similarly exploit
restricted entanglement structure in one-dimensional systems
\citep{Cramer2010}. These approaches illustrate a general principle: scalable
tomography requires a justified restriction on the state family and a
measurement ensemble that identifies the parameters of that family. A
restriction that is too weak offers little computational benefit; one that
is too strong introduces bias.

\subsection{Pure-state likelihood fitting and variational wavefunctions}
When purity is justified, the estimate can be parameterized directly by a
normalized vector $|\psi_\theta\rangle$ rather than a general density matrix.
The likelihood for independent outcomes $\{(y_i,\Pi_i)\}_{i=1}^M$ is
\begin{equation}
 \mathcal P(\mathcal D|\theta)
 =\prod_{i=1}^M
 \operatorname{Tr}\!\left(
 |\psi_\theta\rangle\langle\psi_\theta|\Pi_i
 \right),
 \qquad
 \widehat\theta=\arg\max_\theta\mathcal P(\mathcal D|\theta).
 \label{eq:background-mle}
\end{equation}
Equivalently, one minimizes the negative log likelihood. The pure-state
constraint reduces the number of degrees of freedom relative to a general
mixed-state model, but makes the optimization nonconvex. A direct complex
vector and a neural function are two alternative coordinate systems for
the same physical pure-state manifold.

Neural quantum states use a neural network to map a computational-basis
configuration $x$ to a complex amplitude $\psi_\theta(x)$. The original
neural quantum-state proposal demonstrated that this representation can
capture nontrivial many-body correlations in selected physical systems
\citep{Carleo2017}. In tomography, the same idea can be trained against
measurement probabilities rather than an energy expectation
\citep{Torlai2018}. Experimental work has also combined neural wavefunctions
with models of known measurement errors
\citep{Torlai2019}. These developments motivate neural variational tomography,
but they do not imply that every neural architecture is efficient for every
state family. A network may be expressive enough to represent a state while
the likelihood landscape remains difficult to optimize from a particular
initialization.

Neural tomography also encompasses models that do not directly parameterize
a pure wavefunction. Torlai and Melko introduced neural density operators
through latent-space purification \citep{TorlaiMelko2018}.
Carrasquilla \emph{et al.} learned informationally complete measurement-outcome
distributions with generative models without assuming purity
\citep{Carrasquilla2019}. Ahmed \emph{et al.} used conditional generative
adversarial networks with a physical density-matrix output for optical-state
tomography \citep{Ahmed2021}. More recently, Li \emph{et al.} demonstrated
recurrent generative-model reconstruction of GHZ states on a superconducting
processor with up to five qubits \citep{Li2025}. These studies establish
alternative neural representations and experimental precedents; the present
work instead isolates a phase-conditioned parameterization within ideal
pure-state likelihood fitting.

Representational results likewise depend on the target family. Restricted
Boltzmann machines can compactly represent selected states with volume-law
entanglement \citep{Deng2017}; this is not a statement that generic Haar
states admit comparable compression. Autoregressive recurrent wavefunctions
permit direct, independent sampling in their defining basis
\citep{HibatAllah2020}. These capabilities motivate structured models while
leaving cross-basis probability evaluation and tomography optimization as
separate questions.

It is useful to distinguish three notions that are sometimes conflated:
(i) \emph{representation capacity}, meaning that some parameter value
describes the target; (ii) \emph{statistical identifiability}, meaning that
the measurement distributions distinguish the target from alternatives;
and (iii) \emph{optimization attainability}, meaning that the chosen
training procedure finds a good parameter value from its initialization.
The GHZ phase failure examined here is primarily a diagnostic of the third
notion, in a regime where the second is limited by the frequency of
phase-sensitive settings. It is not evidence that the neural model cannot
represent the target.

\subsection{Classical shadows and why the postprocessing rule matters}
Classical-shadow protocols construct an estimator from randomized
measurements and use it to predict many observables without reconstructing
the full density matrix for every query \citep{Huang2020}. For local Pauli
measurements, each qubit's inverse measurement channel maps the observed
projector to $3|b,z\rangle\langle b,z|-I$. Tensoring these factors yields
an unbiased snapshot, but averaging the snapshots does not preserve
positive semidefiniteness. A physical density matrix can be obtained by
projecting onto the positive trace-one set, but Euclidean projection,
eigenvalue clipping followed by renormalization, and leading-eigenvector
truncation are distinct estimators. Their differences are not a minor
implementation detail: they impose different spectral biases and can change
fidelity substantially.

Projected least-squares tomography has also been analyzed directly:
Gu\c{t}\u{a} \emph{et al.} established nonasymptotic trace-distance guarantees
for specified measurement ensembles \citep{Guta2020}. Those guarantees
illustrate that a projection-based estimator should be assessed together
with its measurement model, rather than equated with arbitrary eigenvalue
clipping. Experimental estimation of state properties from classical
shadows has been demonstrated by Struchalin \emph{et al.}
\citep{Struchalin2021}. A related but distinct task is direct fidelity
estimation, which estimates overlap with a specified target from Pauli
measurements without reconstructing an arbitrary unknown state
\citep{FlammiaLiu2011}.

The shadow representation is particularly useful for observable prediction,
whereas full-state reconstruction asks for a single physical matrix or
wavefunction. The optimal estimator depends on the task and loss function.
The full-matrix Frobenius variance and the variance of a selected Pauli
observable are also different quantities; one should not use the latter as
a surrogate for the former. We therefore include shadow estimators as
explicitly defined baselines, but compare their final physical states with
the likelihood estimators only after specifying the projection rule.

\subsection{Why a structured phase prior is a testable inductive bias}
A GHZ state is a coherent superposition of macroscopically distinct
computational-basis configurations. Its defining coherence is carried by
off-diagonal density-matrix elements
$\rho_{0^N,1^N}=e^{-i\varphi}/2$, not by the populations alone.
The phase $\varphi$ is observable through multi-qubit correlations, which
is the central feature of GHZ-type nonclassicality
\citep{Greenberger1989,Mermin1990}. Under local Pauli measurements, the
all-$X/Y$ subset directly probes this coherence, as derived in
Section~\ref{sec:problem}. When bases are selected uniformly and
independently, the fraction of such shots decays exponentially in $N$.

A measurement-derived phase anchor is therefore a form of
data-conditioned inductive bias. It is not an oracle: the anchor is
estimated from observed outcomes and can be wrong when there are too few
phase-sensitive records or when the assumed GHZ structure is violated.
The bounded residual phase head lets the model refine the anchor without
freely rediscovering an arbitrary phase branch. The preselection rule is
correspondingly treated as a hypothesis test for a narrow canonical state
family, not as a general entanglement witness. This framing connects the
architectural modification to a physically interpretable feature of
the measurement channel.

Measurement alignment can also act on the acquisition protocol itself.
Adaptive single-qubit tomography uses preliminary observations to choose
subsequent measurement bases \citep{Mahler2013}. For specified Pauli
observables, derandomization replaces random settings with a deterministic
measurement schedule \citep{Huang2021}, while locally biased shadows tune
basis probabilities to the Hamiltonian being estimated
\citep{Hadfield2022}. These approaches change how data are collected. Our
phase-aware procedure is complementary: it uses an existing uniformly
random local-Pauli record to condition the state representation, without
claiming to implement or benchmark an adaptive measurement protocol.

\subsection{Scope relative to information-theoretic limits}
The statistical limits depend on the measurement model and error metric.
Optimal estimation of an unknown Haar-distributed pure state gives an
ensemble-mean fidelity bound \citep{Bruss1999,Hayashi2005}.
For general density operators, sample-optimal tomography and efficient
collective-measurement constructions have been developed by Haah
\emph{et al.} \citep{Haah2017} and O'Donnell and Wright
\citep{ODonnellWright2016}. Restrictions to single-copy measurements lead
to distinct lower-bound questions, including those analyzed by Lowe and
Nayak \citep{LoweNayak2025}. These results address different statistical
criteria and measurement resources; they do not directly predict the
fidelity of the finite GHZ experiments reported here. They do, however, reinforce an essential
distinction: an estimator can improve numerical efficiency or use a
structural prior, but cannot reconstruct information that is absent from
the measurement record. The present work makes no claim to evade such
information-theoretic limits.

\section{Mathematical formulation and estimators}
\label{sec:problem}

\subsection{State space, measurement operators, and identifiability}
We consider an $N$-qubit system with Hilbert space
$\mathcal H=(\mathbb C^2)^{\otimes N}$ and dimension $d=2^N$.
A general physical state is a density operator
$\rho\in\mathcal D_d=\{\rho\succeq0:\operatorname{Tr}\rho=1\}$.
A pure state has the rank-one form
$\rho=|\psi\rangle\langle\psi|$, where
$\langle\psi|\psi\rangle=1$ and the vectors $|\psi\rangle$ and
$e^{i\gamma}|\psi\rangle$ describe the same physical state. The pure-state
manifold therefore has real dimension $2d-2$; the full-rank density-operator
space has dimension $d^2-1$. In this paper purity is an explicit modeling
assumption. It is not inferred or certified from the measurement record.

A measurement is described by positive operators
$\{\Pi_y\}_y$ satisfying $\sum_y\Pi_y=I$. The Born rule gives
$p_\rho(y)=\operatorname{Tr}(\rho\Pi_y)$
\citep{Hradil1997,ParisRehacek2004}. For local Pauli measurements, a setting
$\mathbf b=(b_1,\ldots,b_N)$ specifies one of the eigenbases of
$X,Y,Z$ on each qubit, and an outcome
$z=(z_1,\ldots,z_N)\in\{0,1\}^N$ selects one rank-one product projector
\begin{equation}
 \Pi_{z|\mathbf b}=
 \bigotimes_{j=1}^N
 |b_j,z_j\rangle\langle b_j,z_j|.
 \label{eq:qst-projector}
\end{equation}
Equivalently, with $U_X=H$, $U_Y=HS^\dagger$, and $U_Z=I$,
$|b_j,z_j\rangle=U_{b_j}^\dagger|z_j\rangle$. Thus
\begin{equation}
 p_\rho(z|\mathbf b)=
 \operatorname{Tr}(\rho\Pi_{z|\mathbf b})
 =|\langle z|U_{\mathbf b}|\psi\rangle|^2,
 \quad U_{\mathbf b}=\bigotimes_{j=1}^N U_{b_j}.
 \label{eq:qst-born}
\end{equation}
The total copy budget $M$ counts recorded outcomes, not distinct settings:
one copy is consumed for each pair $(\mathbf b,z)$.

The complete family of product Pauli measurements is informationally
complete. To see why, expand
\begin{equation}
 \rho=\frac1d\sum_{P\in\{I,X,Y,Z\}^{\otimes N}}
 r_P P,\qquad r_P=\operatorname{Tr}(\rho P).
 \label{eq:pauli-expansion}
\end{equation}
For each Pauli string $P$, choose a local basis that matches its nonidentity
factor at every site; the product of the corresponding single-qubit
outcomes estimates $r_P$. Since the Pauli strings form an orthogonal
operator basis, the full set of these expectation values determines $\rho$.
Informational completeness is an identifiability statement in the
infinite-data limit. It does not imply that every coefficient is estimated
with comparable variance under a finite random allocation of settings.

For a physical estimate $\sigma$, we report squared fidelity
$F(\rho,\sigma)=(\operatorname{Tr}\sqrt{\sqrt\rho\,\sigma\sqrt\rho})^2$.
For a pure target this reduces to
$F=\langle\psi|\sigma|\psi\rangle$. An overlap with a nonpositive matrix
is not reported as a state fidelity.

\subsection{Forward likelihood with vector and neural parameterizations}
For each distinct observed event $k=(\mathbf b_k,z_k)$, let $c_k$ be its count,
$\sum_k c_k=M$, and $v_k=\langle z_k|U_{\mathbf b_k}$.
For a normalized candidate $\psi_\theta$, the unregularized objective is
\begin{equation}
 L(\theta)=-\frac1M\sum_k c_k
       \log\left|v_k\psi_\theta\right|^2.
 \label{eq:qst-nll}
\end{equation}
We compare a directly parameterized complex vector and a neural function
$f_\theta:\{0,1\}^N\to\mathbb C$. In either case the vector used in
Eq.~\eqref{eq:qst-nll} is normalized exactly:
\begin{equation}
 \psi_\theta(x)=\frac{f_\theta(x)}{
     \sqrt{\sum_{x'}|f_\theta(x')|^2}}.
 \label{eq:qst-normalization}
\end{equation}
The resulting $\sigma_\theta=|\psi_\theta\rangle\langle\psi_\theta|$ is
Hermitian, positive semidefinite, trace one, and rank one whenever the
unnormalized vector is nonzero. These guarantees are shared by both
parameterizations. Neural wavefunctions and their use in likelihood-based
tomography are established approaches \citep{Carleo2017,Torlai2018}; the present
comparison tests their numerical behavior under a matched measurement protocol.

The direct vector baseline is optimized in real and imaginary coordinates
subject to normalization, while the neural estimator represents the
unnormalized amplitudes by a shared network. Both are pure-state MLE
parameterizations and both have a nonconvex optimization landscape because
the rank-one constraint is nonlinear. This comparison is deliberately
different from comparing a neural pure state with an unconstrained linear
inversion: the latter changes both the state prior and the estimator.
The global phase of $f_\theta$ is unidentifiable and is removed by the
projector $|\psi_\theta\rangle\langle\psi_\theta|$.

For fixed parameters, the population version of Eq.~\eqref{eq:qst-nll} is
\begin{equation}
 L_\infty(\theta)=\overline H(p_\psi)
 +\frac1{3^N}\sum_{\mathbf b}
 D_{\mathrm{KL}}\!\left(p_\psi(\cdot\mid\mathbf b)
                  \Vert p_\theta(\cdot\mid\mathbf b)\right),
 \label{eq:qst-kl}
\end{equation}
where $\overline H$ is the average conditional measurement entropy.
Informational completeness identifies a state when all these distributions
match. It does not ensure representability by a fixed-width neural model,
finite-sample accuracy, or convergence of a nonconvex optimizer. A positive
additive probability stabilizer, if used computationally, modifies this exact
identity: $p_\theta+\varepsilon$ corresponds to the normalized distribution
$(p_\theta+\varepsilon)/(1+d\varepsilon)$ and an additive constant in the
objective. Numerical safeguards are therefore specified separately from the
population theorem.

\subsection{Why GHZ relative phase is measured by equatorial Pauli settings}
The GHZ family considered here is
\begin{equation}
 |\mathrm{GHZ}_\varphi\rangle
 =\frac{|0^N\rangle+e^{i\varphi}|1^N\rangle}{\sqrt2},
 \qquad \varphi\in[-\pi,\pi).
 \label{eq:qst-ghz-family}
\end{equation}
Its density operator contains two diagonal terms and two coherence terms:
\begin{align}
 \rho_\varphi=\frac12\big(&|0^N\rangle\langle0^N|
 +|1^N\rangle\langle1^N|\\
 &+e^{-i\varphi}|0^N\rangle\langle1^N|
 +e^{i\varphi}|1^N\rangle\langle0^N|\big).
 \label{eq:qst-ghz-density}
\end{align}
A local Pauli string containing at least one $Z$ has zero matrix element
between $|0^N\rangle$ and $|1^N\rangle$, so its expectation does not
depend on $\varphi$. In contrast, let
$P_{\mathbf b}=\bigotimes_j b_j$ for an all-$X/Y$ setting and let $k$ be
the number of $Y$ factors. Since
$\langle0|X|1\rangle=1$ and
$\langle0|Y|1\rangle=-i$,
\begin{equation}
 \langle P_{\mathbf b}\rangle_{\varphi}
 =\cos\!\left(\varphi-\frac{k\pi}{2}\right).
 \label{eq:qst-ghz-correlator}
\end{equation}
For a complete outcome string $z$, define its parity
$s(z)=(-1)^{\sum_j z_j}$. The all-$X/Y$ outcome distribution is
\begin{equation}
 p_\varphi(z|\mathbf b)
 =2^{-N}\left[
 1+s(z)\cos\!\left(\varphi-\frac{k\pi}{2}\right)
 \right].
 \label{eq:qst-ghz-outcome}
\end{equation}
Thus all-$X/Y$ records carry direct information about the relative phase,
whereas a setting with at least one $Z$ axis has a phase-independent
distribution for the ideal two-component GHZ family. This statement is
specific to the ideal canonical GHZ family; for a general wavefunction,
settings containing $Z$ can still constrain off-support amplitudes and
other coherences.

Under independent uniform local basis selection,
\begin{equation}
 q_{XY}=\Pr(\mathbf b\in\{X,Y\}^N)
 =\left(\frac23\right)^N,\qquad
 \mathbb E[M_{XY}]=M\left(\frac23\right)^N.
 \label{eq:qst-ghz-shot-count}
\end{equation}
For $N=10$, this predicts about 69 all-$X/Y$ shots at $M=4,000$ and
138 at $M=8,000$. The two same-target records used in the budget
verification contain 65 and 122 such shots, respectively.

The phase-only negative log likelihood, up to a phase-independent constant,
is
\begin{equation}
 \mathcal L_{\mathrm{GHZ}}(\varphi)
 =-\sum_{i\in\mathcal I_{XY}}
 \log\left[
 1+s_i\cos\!\left(\varphi-\frac{k_i\pi}{2}\right)
 \right],
 \label{eq:qst-ghz-phase-nll}
\end{equation}
where $\mathcal I_{XY}$ indexes all-$X/Y$ observations and $k_i$ is the
number of $Y$ axes in shot $i$. This one-dimensional likelihood motivates
the measurement-derived phase anchor used in the exploratory architecture
of Section~\ref{sec:phaseaware}.

Fisher information connects a measurement distribution to local statistical
distinguishability, as developed in quantum estimation by Braunstein and
Caves \citep{BraunsteinCaves1994}. Specializing this viewpoint to
Eq.~\eqref{eq:qst-ghz-outcome}, for a generic phase away from singular
boundary points, each all-$X/Y$ shot contributes one unit of classical
Fisher information for $\varphi$.
Consequently the phase Fisher information in this ideal restricted model is
\begin{equation}
 \mathcal I_\varphi=M_{XY},\qquad
 \operatorname{Var}(\widehat\varphi)\gtrsim \frac{1}{M_{XY}},
 \label{eq:qst-ghz-fisher}
\end{equation}
where the variance expression is the local Cram\'er--Rao heuristic for an
efficient unbiased estimator, not a finite-sample guarantee for the neural
tomography problem. Since $M_{XY}$ is typically of order
$M(2/3)^N$, the direct phase information becomes sparse as $N$ grows.
This provides a measurement-theoretic explanation for why phase learning
may be less stable than population learning under the chosen random-basis
protocol.

\subsection{Local-shadow reconstruction and spectral postprocessing}
For a recorded shot, the local classical-shadow snapshot is
\begin{equation}
 \hat\rho^{(m)}=\bigotimes_{j=1}^N
 \left(3U_{b_j^{(m)}}^\dagger|z_j^{(m)}\rangle
       \langle z_j^{(m)}|U_{b_j^{(m)}}-I_2\right),
 \qquad
 A=\frac1M\sum_m\hat\rho^{(m)}.
 \label{eq:qst-shadow}
\end{equation}
This is an unbiased estimator of $\rho$, although $A$ need not be positive
\citep{Huang2020}. We distinguish three postprocessing operations: Euclidean
projection onto the density matrices, clipping negative eigenvalues followed
by proportional renormalization as a diagnostic, and the projector onto a
leading eigenvector as a rank-one spectral baseline. These are different
estimators. In particular, if $A=Q\operatorname{diag}(\mu_i)Q^\dagger$,
the Euclidean projection is
\begin{equation}
 \Pi_{\mathcal D}(A)=Q\operatorname{diag}
       \bigl(\max(\mu_i-\tau,0)\bigr)Q^\dagger,
 \qquad \sum_i\max(\mu_i-\tau,0)=1.
 \label{eq:qst-simplex}
\end{equation}
The common shift $\tau$ is essential \citep{Smolin2012}. This least-squares
projection is not generally the multinomial maximum-likelihood estimator.

The full Frobenius variance of Eq.~\eqref{eq:qst-shadow} can be calculated
without an asymptotic approximation. A single-qubit factor has eigenvalues
$2,-1$, and hence squared Frobenius norm $5$. Independence of the shots and
unbiasedness give
\begin{equation}
 \mathbb E\|A-\rho\|_F^2
   =\frac{5^N-\operatorname{Tr}(\rho^2)}{M}.
 \label{eq:qst-shadow-mse}
\end{equation}
The $3^w$ second moment associated with a weight-$w$ Pauli observable must
not be substituted for this full-matrix error. A poor result for a particular
postprocessing rule is also not a lower bound on all reconstruction algorithms.

\subsection{A limited guarantee for residual symmetry projection}
For residual denoising, define the real vector space
$\mathcal V_0=\{D:D=D^\dagger,\operatorname{Tr}D=0\}$ and equip complex
matrices with the real inner product
$(A,B)=\Re\!\left[\mathrm{Tr}\!\left(A^{\dagger} B\right)\right]$.
For $M_H=(M+M^\dagger)/2$, the map
\begin{equation}
 \Pi_0(M)=M_H-\frac{\operatorname{Tr}M_H}{d}I
 \label{eq:qst-residual-projection}
\end{equation}
is the orthogonal projection onto $\mathcal V_0$. For any true residual
$D\in\mathcal V_0$,
\begin{align}
 \|\Pi_0(M)-D\|_F^2
 ={}&\|M-D\|_F^2
 -\left\|\frac{M-M^\dagger}{2}\right\|_F^2
 -\frac{|\operatorname{Tr}M_H|^2}{d}.
 \label{eq:qst-projection-identity}
\end{align}
This identity follows by decomposing $M-D$ into traceless Hermitian,
real-identity, and skew-Hermitian components. It guarantees non-increase of
the residual error, with equality when no component is removed. It does not
guarantee positive semidefiniteness of a subsequent state update, nor does it
establish stability or convergence of a learned iteration. For this reason,
the denoiser study reports symmetry and residual errors separately and uses a
common physical-state projection before comparing state fidelities. Results
for a particular real convolutional network are not a no-go theorem for real
or complex neural networks; two real channels are a lossless representation
of a complex input.

\section{Resource accounting and statistical scope}
\label{sec:scope}

\subsection{Parameter storage and working memory}
A two-hidden-layer fully connected network with width $H$, $N$ real inputs,
and two real outputs has
\begin{equation}
 P(N,H)=HN+H^2+4H+2
 \label{eq:qst-parameters}
\end{equation}
parameters. This is linear in $N$ when $H$ is fixed. It describes the stored
model, not the working memory of exact likelihood evaluation, and keeping $H$
fixed is not a guarantee of fixed approximation accuracy. A general pure
state has $2d-2$ real degrees of freedom.

A counting argument makes the fixed-precision limitation explicit. Fix a
decoder and include all stored state-dependent information and any architecture
description in a budget of $B$ bits. There are at most $2^B$ decoded pure states.
For a fixed candidate $|\phi\rangle$ and a Haar target $|\psi\rangle$, unitary
invariance gives the overlap $t=|\langle\phi|\psi\rangle|^2$ the density
$(d-1)(1-t)^{d-2}$ on $[0,1]$. Hence, for $0<\epsilon<1$,
\begin{equation}
 \begin{aligned}
 \Pr_\psi\!\left[|\langle\phi|\psi\rangle|^2\geq1-\epsilon\right]
 &=\epsilon^{d-1},\\
 \Pr_\psi\!\left[\max_{\phi\ \mathrm{encoded}}
       |\langle\phi|\psi\rangle|^2\geq1-\epsilon\right]
 &\leq\min\{1,2^B\epsilon^{d-1}\}.
 \end{aligned}
 \label{eq:qst-encoding-union}
\end{equation}
The second inequality is the union bound. Covering a fraction at least
$1-\delta$ of Haar targets, with $0\leq\delta<1$, therefore requires
\begin{equation}
 B\geq(d-1)\log_2(1/\epsilon)+\log_2(1-\delta).
 \label{eq:qst-encoding-bits}
\end{equation}
At fixed accuracy and nonzero covered fraction, this requires $\Omega(2^N)$
bits. Thus $O(N)$ parameters at fixed precision cannot provide generic
high-fidelity Haar-state compression. This is a representation bound, not
a runtime theorem, and it does not preclude compact descriptions of
restricted structured families.

Exact normalization in Eq.~\eqref{eq:qst-normalization} evaluates all $d$
amplitudes. If the $K$ measurement rows are stored densely, they require
$O(Kd)$ memory, while batched neural activations require $O(LHd)$ memory for
$L$ hidden layers. The computational work also includes the cost of the
network evaluations, approximately
$O(d[NH+(L-1)H^2]+Kd)$ for dense overlaps.
The sparse implementation exploits the zeros imposed by $Z$ measurements:
a row with $k_{XY}$ non-$Z$ axes has $2^{k_{XY}}$ nonzero entries, and
an independently uniform basis choice obeys
\begin{equation}
 \mathbb E[2^{k_{XY}}]=(5/3)^N.
 \label{eq:qst-sparse-rows}
\end{equation}
Thus, before duplicate-event aggregation, the expected total number of row
entries is $M(5/3)^N$; aggregation can only reduce it. This improves the expected
row-storage exponent relative to dense storage but retains exponential scaling.
Autoregressive normalization alone does not remove the
coherent sum needed for a generic rotated-basis amplitude.

We therefore distinguish model bytes, measured working memory, and runtime.
The relevant representation comparison for pure targets is a complex vector
of $d$ amplitudes; avoiding a dense $d\times d$ matrix is already possible
without a neural network. Formula-based memory extrapolations to larger $N$
are not reported as successful reconstructions at those sizes.

\subsection{Information limits and interpretation of fidelity}
For an independently Haar-distributed unknown pure target in dimension $d$,
an estimator using only $M$ copies obeys the optimal mean-fidelity bound
\begin{equation}
 \mathbb E[F]\leq\frac{M+1}{M+d},
 \label{eq:qst-haar-bound}
\end{equation}
even if collective measurements are allowed
\citep{Bruss1999,Hayashi2005}. The expectation includes target and measurement
randomness; Eq.~\eqref{eq:qst-haar-bound} is not a deterministic upper bound
on every individual run. Ground-truth amplitudes and fidelities are available
to the simulator for generation and final evaluation only. Any use of them
for parameter fitting, model selection, or stopping would invalidate the
interpretation as reconstruction of an unknown state from the stated records.

Avoiding inverse-channel arithmetic also does not remove all statistical
limitations of local measurements. Consider
$|\mathrm{GHZ}_{\pm}\rangle=(|0^N\rangle\pm|1^N\rangle)/\sqrt2$.
Only all-$X/Y$ settings with an even number of $Y$ axes distinguish their
signs. Such a basis occurs with probability $q=2^{N-1}/3^N$; all other
settings have identical outcome distributions for the two states. With equal
priors, the optimal discrimination error after $M$ shots is
\begin{equation}
 P_{\mathrm{err}}^\star=\frac12(1-q)^M.
 \label{eq:qst-ghz-bound}
\end{equation}
Thus a restricted compact state family can still require exponentially many
uniformly random local measurements for a particular global feature.
Equation~\eqref{eq:qst-ghz-bound} is an exact result for the two-sign problem,
not a quantitative bound for the continuous random-phase GHZ benchmark below.
Forward likelihood can be statistically preferable to a chosen inverse
estimator without circumventing this lack of information.

The experiments are simulations of ideal pure-state data. They do not
establish tomography of noisy mixed states, recovery of a mixed state's
principal eigenvector by pure-state maximum likelihood, or error mitigation.
Their scope is the specified state families, measurement budgets,
parameterizations, and optimization schedules.

\section{Materials and Methods}
\label{sec:experiments}
\subsection{Targets, copy budgets, and independent random streams}
The main numerical comparison uses five Haar conditions with five independent
target/data cases per condition, and a separate GHZ cohort with ten cases at
each of $N=4,8,10$. Haar states are generated by normalizing independent
complex-Gaussian amplitude vectors. Their system-size and copy-budget pairs are
\begin{equation}
\begin{aligned}
(N,M)\in\{&(4,15000),\ (8,8000),\ (8,16000),\\
          &(10,8000),\ (10,32000)\}.
\end{aligned}
\end{equation}
The GHZ targets have the form
\begin{equation}
 |\mathrm{GHZ}_\varphi\rangle=
 \frac{|0^N\rangle+e^{i\varphi}|1^N\rangle}{\sqrt2},
 \qquad \varphi\sim\mathrm{Uniform}[-\pi,\pi),
\end{equation}
and every method receives the same 8,000 recorded outcomes within a case.
Randomizing the phase avoids supplying a known positive GHZ phase. The
unrestricted vector and generic neural model are not restricted to the
two-dimensional GHZ support. Earlier lower-budget GHZ records are archived
separately and are not pooled with these main-table estimates.

For each case, independent child streams generate the target, training
measurements, held-out measurements, and optimization initializations. Every
training shot uses an independently uniform local basis and one Born-rule
outcome. The simulator groups repeated bases only to reuse an exact probability
calculation; outcomes remain independent. Every estimator receives the same
fixed $M$ training records. An additional 2,000 independent shots per target
are retained solely for held-out likelihood evaluation. They are not used
for parameter fitting, restart selection, scheduling, or stopping and are
reported separately from the training-copy budget. Exact target amplitudes
are used only for simulation and final diagnostic evaluation.

The complete per-shot bases, bitstrings, target vectors, random seeds, selected
fit parameters, objective traces, and machine-readable summaries are retained.
SHA-256 digests link the raw records to each result. The analysis checks record
counts, matching configurations, distinct target seeds, saved-state normalization,
recomputed fidelities, and training-likelihood restart selection before producing
any publication table. No target is excluded by fidelity; unsuccessful candidate fits remain in
the archive even when a lower-loss candidate is selected.

\subsection{Neural models and matched likelihood optimization}
The neural model is a real fully connected network with two hidden GELU layers,
zero-centered spin inputs $2x-1$, and two output heads. Haar experiments use
Cartesian outputs $f_\theta(x)=y_0(x)+iy_1(x)$ with width $H=64$ at four qubits
and $H=256$ at eight and ten qubits. GHZ experiments use
\begin{equation}
 f_\theta(x)=\exp(s_\theta(x)+i\phi_\theta(x)),\qquad
 s_\theta=\mathrm{clip}(y_0,-15,5),\quad \phi_\theta=y_1,
\end{equation}
with $H=64$ at four qubits and $H=128$ at eight and ten qubits.
Clipping is a numerical safeguard, not an exact sparsity constraint. All
$d$ amplitudes are evaluated and normalized; neither model avoids that
exponential enumeration in this implementation.

Hidden-layer weights are independent zero-mean Gaussians with standard
deviation $\sqrt{2/n_{\mathrm{in}}}$; output weights have standard deviation
$1/\sqrt H$, and all biases start at zero. The unrestricted vector fit starts
from the normalized output of the corresponding initialized neural model.
Thus the vector and neural methods start from exactly the same physical state
for each of two initialization seeds, while using different optimization
coordinates. The original two seeds were fixed before the primary experiments.
The subsequent multistart extensions are specified below.

Both methods use the same optimizer sequence: 800 Adam steps for $N=4,8$ or
1,600 for $N=10$ \citep{KingmaBa2015}, followed by at most 600 L-BFGS-B
iterations \citep{Byrd1995}. Adam uses
$\beta_1=0.9$, $\beta_2=0.999$, denominator stabilizer $10^{-8}$, and a cosine
learning-rate schedule from 0.005 to 0.0005. L-BFGS-B uses history size 10,
gradient tolerance $10^{-7}$, relative-function tolerance $10^{-11}$, and a
40-trial line-search limit, with its function-evaluation limit set to 3,700.
The final restart is selected solely by the smaller
training NLL. Equal iteration limits do not imply equal arithmetic or runtime;
both are reported rather than assuming equal computational cost.

In finite precision the optimized conditional distribution is
\begin{equation}
 q_\theta(z\mid\mathbf b)=
 \frac{p_\theta(z\mid\mathbf b)+10^{-12}}{1+d\,10^{-12}}.
\end{equation}
It is normalized in every basis. The additive constant in its NLL is included,
and analytic gradients include the derivative of exact wavefunction
normalization. Finite-difference tests independently check the complex
amplitude gradients, Cartesian and polar backpropagation, and convolutional
gradients. Further tests verify exact population stationarity and recovery
from noiseless informationally complete two-qubit probabilities.

Development-only pilots were used to identify poor initialization and inadequate
optimization progress through training likelihood. Those targets and records
are excluded from the primary statistics. The original protocol was frozen before
the primary cases were started. Later initialization checks are labeled as
follow-up analyses, not part of that original protocol. The results concern this bounded optimization
procedure; neither an optimizer status flag nor a small objective gap certifies
a global optimum.

\subsection{Uniform 8,000-shot GHZ comparison}
\label{sec:uniform8000-protocol}
To compare all estimators at a common larger measurement budget, we generate
an independent follow-up cohort at $N=4,8,10$, with ten random-phase GHZ
targets and independently generated measurement records at each size.
Every case contains exactly 8,000 training shots, shared by all six methods:
shadow clipping, Euclidean simplex projection, rank-one shadow projection,
unrestricted statevector likelihood fitting, generic polar neural fitting,
and phase-aware neural fitting. An additional 2,000 independent shots per
case are evaluation-only; they do not enter fitting, phase anchoring,
restart selection, or stopping. No target, record, or restart is removed
on the basis of fidelity. These new records are neither extensions of nor
replacements for any historical raw measurement record.

The protocol and random seeds were fixed before generating or inspecting
this cohort. A master seed of 731984521 and separate child streams specify
the target phase, training record, evaluation record, and initializations.
The initial fixed-protocol comparison used two starts per fitted method,
with width 64 at four qubits and 128 at eight and ten qubits for both neural models. All three likelihood methods use
800 full-batch Adam steps at $N=4,8$, or 1,600 at $N=10$, followed by up to
600 L-BFGS-B iterations, with the same learning-rate schedule, tolerances,
and history size as the original benchmark. This matched the original optimizer-step budgets across fitted models
within each system size, not their arithmetic cost or wall-clock time.
The present main table additionally strengthens the statevector baseline
with the post hoc initialization audit in Section~\ref{sec:sv-robustness};
the neural protocols are unchanged. The statevector starts from the exact normalized
output of the generic network at the same initialization seed. The
phase-aware network shares the generic network's hidden and amplitude
initialization, but uses its own measurement-derived phase anchor and
small residual-phase initialization. The different initial phases and
restricted phase family are part of the combined method, not controlled
away by the equal parameter counts.

The phase-aware branch uses the previously fixed screen
$S_Z\geq0.85$, $M_{XY}\geq20$, residual bound $\delta_{\max}=0.30$, and
phase-head initialization scale 0.02. A rejection would route the case to
the generic estimator rather than remove it. In the original comparison, each method selected the candidate with
minimum training NLL across its two starts. The strengthened statevector
comparison uses four candidates under the same training-only criterion. The
selection file is written before the evaluation target or held-out outcomes
are opened. All 180 fitted parameter vectors and objective histories are
retained, including capped or unsuccessful optimizer terminations.

As a structured control, we also evaluate the direct phase-only estimate
\begin{equation}
 |\widehat\psi_{\mathrm{GHZ}}\rangle
 =\frac{|0^N\rangle+e^{i\widehat\varphi}|1^N\rangle}{\sqrt2},
 \label{eq:direct-ghz-control}
\end{equation}
using the same training-only phase estimate as the anchored network. It adds
no shots and no neural training. This control distinguishes what the ideal
GHZ support and phase prior achieve by themselves from what the flexible
neural refinement contributes; it is reported separately in
Appendix~\ref{app:uniform8000-diagnostics}.

The new cohort uses Python 3.13.5, NumPy 2.3.5, SciPy 1.17.0,
float64/complex128 arithmetic, and three concurrent CPU workers with one
BLAS thread each. Processor identities and run environments are retained in
the reproducibility manifests. Shared data acquisition and likelihood
construction are timed separately. Numerical summaries use the ten
independent target/data cases, not the two nested restarts, as replicates.
Every saved wavefunction is reconstructed from its parameters; training and
held-out likelihoods, fidelities, hashes, and restart choices are recomputed.
Shadow fidelities are additionally checked through an independent spectral
reconstruction of the shot-derived matrix.

\subsection{Multistart selection and optimization checks}
\label{sec:sv-robustness}
\label{sec:selection}
All fitted estimators use the same data-only selection criterion:
\begin{equation}
 \widehat{k}=\mathop{\mathrm{arg\,min}}_{k\in\mathcal K}
 \mathcal L(\theta_k;\mathcal D_{\mathrm{train}}).
 \label{eq:restart-selection}
\end{equation}
Here $\mathcal K$ is the declared candidate set for a method and target;
``best fit'' means the smallest achieved training NLL, not the largest
fidelity to the simulated target. Means and standard deviations use one
selected estimate per target. Every target and candidate summary is retained in the
archive, including capped runs; every newly added fit is saved in full.

For the uniform GHZ comparison, the strengthened statevector procedure uses
four candidates on every record: the two original network-output starts,
one training-shadow principal-eigenvector start, and one independent
complex-Gaussian start. The generic and phase-aware neural procedures each
use two starts, with matched per-start step limits. Thus the statevector
baseline has a larger total optimization budget, explicitly indicated in
Table~\ref{tab:ghz_results}. The spectral initializer also requires shadow
construction; it is not included in the earlier likelihood-only memory
profile. The additional starts were applied uniformly, not only to an
unfavorable target.

For the $N=10$, $M=8000$ Haar condition, a separate audit adds two fresh
paired Cartesian-network initializations to both likelihood estimators,
using the same original per-start schedule. Table~\ref{tab:haar_results}
reports four-start training-likelihood selection for both methods in this
condition; the other Haar rows retain their original two-start protocol.
Further diagnostic tests, excluded from this selection, use spectral and
Gaussian vector starts and continue each saved selected vector and neural
fit for up to 2,400 L-BFGS-B iterations with $\mathrm{ftol}=10^{-15}$ and
$\mathrm{gtol}=10^{-10}$. The additional Haar runs use Python 3.13.5 with NumPy 2.3.5 and SciPy
1.17.0; their hardware and thread settings are recorded separately.
No extra measurements are generated. Selections
are written before evaluation data are loaded. Raw measurement records,
objective histories, iteration counts, and numerical verification results
are retained with the supporting reproducibility materials.

Both parameterizations can have initialization-dependent solutions. Their
individual historical trajectories are therefore treated symmetrically as
optimization diagnostics in the archive, rather than used to define either
method's accuracy ceiling.

\subsection{Shadow estimators and controlled residual denoising}
The three shadow baselines apply clip-and-renormalize, exact simplex projection,
or leading-eigenvector projection to the same unbiased average in
Eq.~\eqref{eq:qst-shadow}. The implementation accumulates Pauli coefficients
before transforming them into a matrix, rather than allocating one dense matrix
per shot. It nevertheless materializes the final $d\times d$ matrix. A raw
shadow's trace and squared Frobenius error are retained as diagnostics;
all reported reconstruction fidelities use physical state estimates.

The separate denoiser study uses four-qubit Haar states and 256 actual Pauli
shots per state. It contains 512 training residual pairs, 128 independent
validation pairs, and eight further test states. For offline pairs,
$\sigma=a\rho+(1-a)\tau$, where $a$ is uniform on $[0,1)$ and $\tau$ is an
independent Haar projector. The input is $A-\sigma$ and the supervised target
is $\rho-\sigma$. No independent Gaussian matrix noise is substituted for the
measurement process.

A three-layer $3\times3$ convolutional network uses channel dimensions
$2\to8\to8\to2$, same padding, GELU hidden activations, and 882 real parameters.
The raw and output-projected networks share initialization, minibatch order,
training examples, and a fixed 60-epoch Adam schedule with batch size 32 and
learning rate 0.001. A third comparison applies the residual projection to
the already trained raw network. Validation results are diagnostic only;
the last fixed epoch is used. Online reconstruction starts from $I/16$ and
performs 40 classical updates with $\eta=0.1$, reusing one fixed 256-shot record.
We compare physical projection only at the end with exact simplex projection
after every update. One training seed is used, so variation over training runs
is not estimated.

\subsection{Statistics, implementation, and resource measurements}
Haar entries are means and sample standard deviations ($\mathrm{ddof}=1$)
over five independent target/data cases; uniform GHZ entries use ten cases
per size. Optimization starts are nested within a case, with one selected
by training NLL, and are not additional replicates. Individual selected
states and paired differences accompany the summaries. Different original
shot-budget conditions use independent target/data cohorts, so pairing
applies to estimator comparisons within a case, not to the comparison
between budgets. The ten-qubit Haar initialization audit reuses all five
original 8,000-shot records unchanged and adds no target replicates.
Historical single-target budget checks remain in the archive and are not
pooled with the main cohorts.

The reference implementation uses Python 3.12.14, NumPy 2.3.5, and SciPy 1.17.0,
with float64/complex128 arithmetic and one BLAS thread per CPU worker.
Two original primary cases could run concurrently. The processor identity,
library versions, and concurrency settings for each execution cohort are
recorded in the reproducibility manifests. Runtime is a measurement of the
reference execution environment, not a hardware-independent performance
guarantee or a timing claim for a particular local computer. Shadow timings include shadow construction
and the specified spectral operation; historical likelihood timings include their original two starts and exclude
shared row construction. Added-fit times and setup costs are recorded
separately; no earlier timing is relabeled as a measurement of the enlarged
candidate bank.

A separate allocation exercise runs each estimator in a fresh process at
$N=4,8,10$. Each likelihood profile uses one restart and 12 L-BFGS-B iterations
to exercise working allocations, not to measure converged accuracy. Operating-
system peak resident memory includes interpreter imports, data simulation,
measurement representations, and optimization or spectral processing. The
profile-only vector initialization is a normalized Gaussian vector to avoid
allocating a neural model; this difference from the main matched initialization
is explicit. Formula-based array sizes through $N=20$ are identified as
analytical extrapolations, with no reconstruction claim at those sizes.

The implementation is CPU-only and does not require a GPU. A portable
local-execution configuration uses one worker and one BLAS thread and
records the processor and software environment for each run. This
configuration does not imply that the reported timings were measured on a
consumer desktop or laptop.

\section{Results}
\label{sec:results}
\subsection{Haar-state reconstruction and copy-budget dependence}
\begin{figure}[tbp]\centering
\manuscriptfigure{0.97\textwidth}{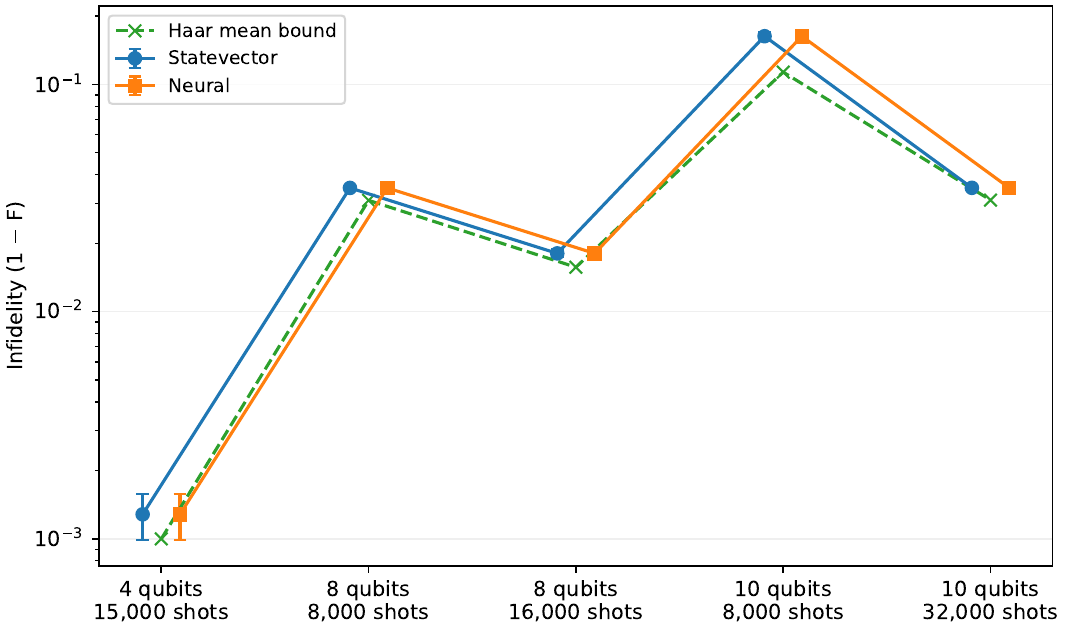}
\caption{Haar-state likelihood reconstruction on the five target/data cases
per condition. Points summarize the same selected estimates as
Table~\ref{tab:haar_results}; error bars show sample standard deviation.
The dashed curve is the optimal Haar ensemble-mean infidelity lower bound,
not a confidence interval for this finite sample.}
\label{fig:fidelity}
\end{figure}

Table~\ref{tab:haar_results} and Figure~\ref{fig:fidelity} compare the physical
estimators on identical data. At $N=8,M=8000$, the vector and neural fits achieve
$F=0.9650\pm0.0005$ and $0.9650\pm0.0005$, respectively. At $N=10,M=8000$, they achieve
$F=0.8369\pm0.0077$ and $0.8369\pm0.0077$. Increasing the ten-qubit budget to 32,000 copies
gives $F=0.9649\pm0.0008$ and $0.9649\pm0.0008$.

These are finite-budget empirical outcomes, not evidence that an estimator
evades information limits. The optimal Haar mean-fidelity bounds are 0.88664
and 0.96902 for the two ten-qubit budgets, respectively. Both the bounds and
the observed dependence on $M$ must be considered when interpreting high
fidelity. The neural fit is compared against the attained vector likelihood
solution, rather than only against a projected inverse estimator.


\begin{table}[tbp]
\centering\small
\setlength{\tabcolsep}{3.5pt}
\caption{Squared reconstruction fidelity for Haar-random pure states.
Entries are means $\pm$ sample standard deviations across five independent
target/data cases per condition. Both likelihood methods use four-start
training-NLL selection at $N=10$, $M=8000$, following the initialization
audit; all other rows retain the original two starts. No selection uses
target fidelity.}
\label{tab:haar_results}
\begin{tabular}{rrccccc}
\toprule
$N$ & $M$ & Clip & Simplex & Rank one & Statevector & Neural \\
\midrule
4 & 15,000 & $0.7622\pm 0.0144$ & $0.9473\pm 0.0135$ & $0.9963\pm 0.0013$ & $0.9987\pm 0.0003$ & $0.9987\pm 0.0003$ \\
8 & 8,000 & $0.0211\pm 0.0009$ & $0.3747\pm 0.0292$ & $0.7865\pm 0.0167$ & $0.9650\pm 0.0005$ & $0.9650\pm 0.0005$ \\
8 & 16,000 & $0.0291\pm 0.0006$ & $0.5206\pm 0.0207$ & $0.8889\pm 0.0030$ & $0.9820\pm 0.0007$ & $0.9820\pm 0.0007$ \\
10 & 8,000 & $0.0025\pm 0.0001$ & $0.0351\pm 0.0076$ & $0.1446\pm 0.0433$ & $0.8369\pm 0.0077$ & $0.8369\pm 0.0077$ \\
10 & 32,000 & $0.0044\pm 0.0001$ & $0.2357\pm 0.0113$ & $0.7003\pm 0.0112$ & $0.9649\pm 0.0008$ & $0.9649\pm 0.0008$ \\
\bottomrule
\end{tabular}
\end{table}

The spectral rule has a large effect on the baseline. At ten qubits and
8,000 copies, clip-and-renormalize gives $F=0.0025\pm0.0001$, whereas the exact
Euclidean projection gives $0.0351\pm0.0076$. Neither operation is the same as the
rank-one spectral estimator or the forward likelihood fit. Consequently,
ratios obtained by dividing neural fidelity by the smallest baseline fidelity
would obscure which methodological change produced the improvement.

\subsection{Uniform-budget reconstruction of structured states}
\label{sec:uniform8000-results}

Table~\ref{tab:ghz_results} reports the selected reconstructions on ten
independent random-phase GHZ records per size, with 8,000 shared shots for
every method. At ten qubits, mean squared fidelities are
$0.9597\pm0.0039$ for the unrestricted vector, $0.9845\pm0.0084$ for the
generic network, and $0.9972\pm0.0030$ for the phase-aware model. At eight
qubits the corresponding means are 0.9858, 0.9923, and 0.9972. The main
table uses the stronger declared candidate bank rather than an isolated
unfavorable start for either representation.

The phase-aware estimate is more accurate than its generic neural
counterpart in all 30 paired cases. Its minimum selected fidelity is
0.9893, and both neural methods have every selected fidelity above 0.9.
The strengthened vector also exceeds 0.9 in every case. These counts
characterize the completed cohort, not a population-level guarantee.
All ten-qubit records contain 117--164 all-$X/Y$ observations within the
fixed 8,000-copy budget, and all records pass the fixed GHZ screen.

The generic and phase-aware networks use equal widths, restart counts, and
per-start optimizer limits. The vector uses four rather than two starts;
its additional initialization cost and all original candidates remain
recorded in the archive. The direct GHZ phase-only estimator is evaluated
on the same records in Appendix~\ref{app:uniform8000-diagnostics}.


\begin{table}[tbp]
\small
\setlength{\tabcolsep}{3pt}

\caption{Uniform 8,000-shot GHZ comparison. Entries are mean squared
fidelity $\pm$ sample standard deviation across the same ten target/data
cases per size. The statevector uses four starts (two network-output,
one spectral, one Gaussian); both neural models use two starts with matched
per-start optimizer limits. Each candidate is selected by training NLL,
never by target fidelity.}
\label{tab:ghz_results}

\centering
\begin{tabular}{@{}rcccccc@{}}
\toprule
$N$
& Clip
& Simplex
& Rank one
& \shortstack{Statevector\\(4 starts)}
& \shortstack{Generic\\neural}
& \shortstack{Phase-aware\\neural} \\
\midrule
4
& $0.6964\pm 0.0130$
& $0.9268\pm 0.0082$
& $0.9952\pm 0.0007$
& $0.9987\pm 0.0002$
& $0.9988\pm 0.0002$
& $0.9995\pm 0.0003$ \\
8
& $0.0214\pm 0.0013$
& $0.4109\pm 0.0427$
& $0.7324\pm 0.0309$
& $0.9858\pm 0.0021$
& $0.9923\pm 0.0026$
& $0.9972\pm 0.0034$ \\
10
& $0.0032\pm 0.0001$
& $0.2551\pm 0.0371$
& $0.4616\pm 0.0243$
& $0.9597\pm 0.0039$
& $0.9845\pm 0.0084$
& $0.9972\pm 0.0030$ \\
\bottomrule
\end{tabular}

\end{table}

\begin{figure}[tbp]\centering
\manuscriptfigure{0.97\textwidth}{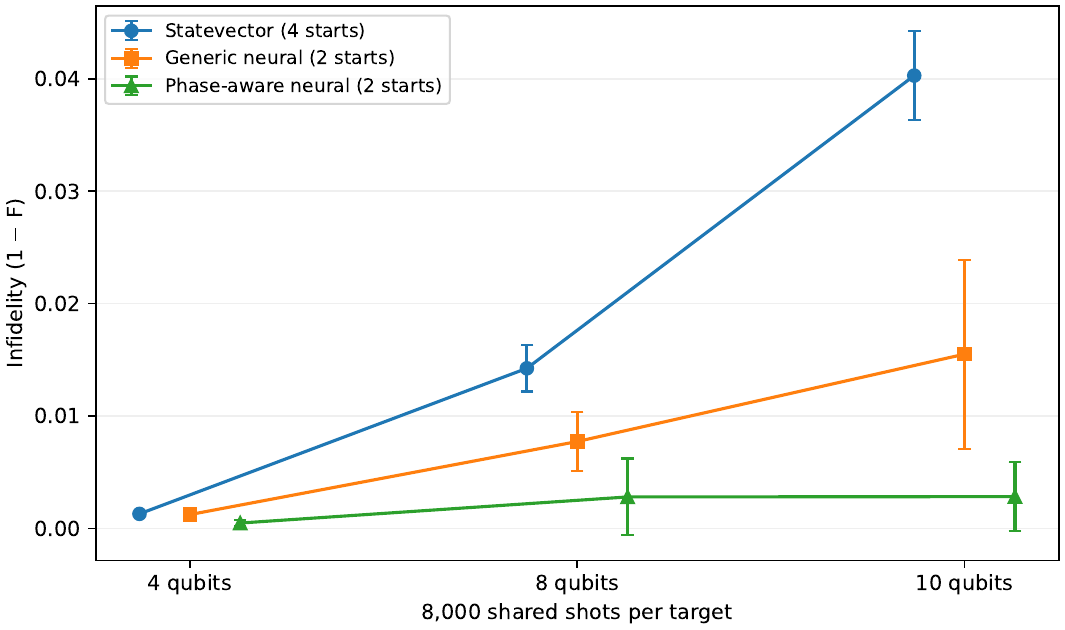}
\caption{Selected likelihood reconstructions on the uniform 8,000-shot GHZ
cohort. Each size contains ten independent target/data cases. The vector
baseline uses four starts, while the neural methods use two starts each;
all selections minimize training NLL. Error bars show sample standard
deviation across all ten records.}
\label{fig:paired}
\end{figure}

\subsection{Optimization stability and information budget}
\label{sec:optimization-summary}
\label{sec:budget-results}
Initialization sensitivity occurs in both vector and neural fits. The main
tables therefore compare the selected states from the stated candidate
banks, not their least favorable starts. The GHZ multistart refinement
improves the unrestricted-vector reference without changing measurements;
the corresponding numerical trajectories for both model types are archived.
This prevents an isolated optimization result from being mistaken for a
limitation of either representation.

The Haar audit tests a different question: whether the ten-qubit,
8,000-copy fidelity can be repaired simply by more starts or tighter local
optimization. Four-start selection gives mean fidelities
$0.8369\pm0.0077$ and $0.8369\pm0.0077$ for vector and neural fitting,
respectively; the individual vector results span 0.8276--0.8435.
The tested alternative starts and continuations do not produce a large
accuracy recovery. This contrasts with the isolated GHZ phase outliers and
is consistent with a substantial finite-data contribution at this sample
budget. It does not prove that the optimizer has found a global optimum or
that every more effective estimator has been excluded. The original,
additional, and diagnostic candidates are all reported in the archive.

The Haar ensemble-mean fidelity bound is 0.88664 at $d=1024$, $M=8000$
(Eq.~\eqref{eq:qst-haar-bound}); it is not a pointwise bound on a five-state
sample. Increasing the copy budget to 32,000 gives the observed mean
fidelity 0.9649 for both likelihood models. The main comparison thus
separates changes in measurement information from changes in optimization
rather than requiring fidelities to follow a predetermined trend.

\subsection{Residual MSE and reconstruction fidelity are distinct}
Across the 128 independent validation residual pairs, the mean residual MSE
per complex matrix entry is 0.00960691 for
the identity input, 0.00246073 for zero output, 0.00197182 for the raw CNN,
0.00196302 after projecting that same CNN output, and 0.00196148 for the
projection-trained CNN. The zero predictor is included because residual
shrinkage can otherwise appear favorable without demonstrating useful
reconstruction. The fixed-output Pythagorean identity is satisfied to a
maximum numerical discrepancy of $2.02\times10^{-16}$.

Despite their lower offline residual MSE, the trained denoisers do not improve
online fidelity in this experiment. With final-only physical projection, the
identity update reaches $0.6362\pm0.0921$, compared with
$0.4578\pm0.0569$ for the raw CNN and $0.4622\pm0.0597$ for the projection-trained
CNN (Table~\ref{tab:denoiser}). This result is limited to the trained architecture,
offline distribution, and update policy. In particular, offline comparison
states have rank at most two, whereas online estimates need not follow that
distribution. The experiment does not constitute a general failure theorem
for real or complex neural denoisers.


\begin{table}[tbp]
\centering\small
\caption{Controlled four-qubit denoiser ablation. All methods reuse the same finite-shot data and update schedule; physical fidelity is evaluated after final density-matrix simplex projection. Validation MSE is averaged over 128 separately generated residual pairs. Reconstruction fidelity is mean $\pm$ sample standard deviation across 8 independent online test targets.}
\label{tab:denoiser}
\begin{tabular}{lrrr}
\toprule
Residual operator & Validation MSE & Final fidelity & Raw Hermiticity error \\
\midrule
Identity & 0.009607 & $0.6362\pm 0.0921$ & $0$ \\
Raw CNN & 0.001972 & $0.4578\pm 0.0569$ & $0.096$ \\
CNN + output projection & 0.001963 & $0.4585\pm 0.0569$ & $0$ \\
Projection-trained CNN & 0.001961 & $0.4622\pm 0.0597$ & $0$ \\
\bottomrule
\end{tabular}
\end{table}

There is a useful exact distinction between the two physical-projection
policies. Let $A_H=(A+A^\dagger)/2$. Euclidean density-matrix projection obeys,
for Hermitian $H$,
\begin{equation}
 \Pi_{\mathcal D}(A)=\Pi_{\mathcal D}(A_H),\qquad
 \Pi_{\mathcal D}(H+cI)=\Pi_{\mathcal D}(H)\quad(c\in\mathbb R).
\end{equation}
The second equality follows because $\operatorname{Tr}\sigma=1$ makes the
change in squared distance under a scalar identity shift independent of the
candidate density matrix. Therefore, for Hermitian $\sigma$ and real $\eta$,
\begin{equation}
 \Pi_{\mathcal D}\bigl(\sigma+\eta\Pi_0(M)\bigr)
 =\Pi_{\mathcal D}\bigl(\sigma+\eta M\bigr).
\end{equation}
The raw CNN and the same CNN with a posthoc residual projection consequently
have identical physical trajectories when simplex projection follows every
update. Their observed maximum fidelity discrepancy is $3.00\times10^{-15}$.
This redundancy does not apply when projection changes the training loss and
learned weights, or when physical projection occurs only at the final step.

\subsection{Working memory and the limits of parameter-count comparisons}
Table~\ref{tab:resources} reports fresh-process allocation profiles. At ten
qubits and 8,000 shots, peak RSS is 134.69 MiB for the vector likelihood
profile, 153.64 MiB for the Cartesian neural profile, and 162.67 MiB
for shadow construction plus simplex projection. The neural parameter vector
alone occupies 0.527 MiB. The likelihood matrix and its adjoint,
rather than those weights alone, account for large working allocations.
These short allocation exercises are separate from the full accuracy study.


\begin{table}[tbp]
\centering\small
\caption{Separate fresh-process allocation profiles. RSS includes interpreter, data acquisition, arrays, and optimizer allocations. Each likelihood profile uses only the configured short L-BFGS allocation exercise; these values do not measure the full main-experiment optimization or establish convergence. All memory units are MiB.}
\label{tab:resources}
\begin{tabular}{rrlrrr}
\toprule
$N$ & $M$ & Estimator & Parameter payload & Sparse $V,V^\dagger$ & Peak RSS \\
\midrule
4 & 15,000 & Statevector & 0.00024 & 0.36 & 59.76 \\
4 & 15,000 & Neural & 0.03517 & 0.36 & 61.27 \\
4 & 15,000 & Simplex & --- & --- & 65.27 \\
8 & 8,000 & Statevector & 0.00391 & 17.96 & 85.82 \\
8 & 8,000 & Neural & 0.52345 & 17.96 & 104.26 \\
8 & 8,000 & Simplex & --- & --- & 69.62 \\
10 & 8,000 & Statevector & 0.01562 & 50.47 & 134.69 \\
10 & 8,000 & Neural & 0.52736 & 50.47 & 153.64 \\
10 & 8,000 & Simplex & --- & --- & 162.67 \\
\bottomrule
\end{tabular}
\end{table}

\begin{figure}[tbp]\centering
\manuscriptfigure{0.99\textwidth}{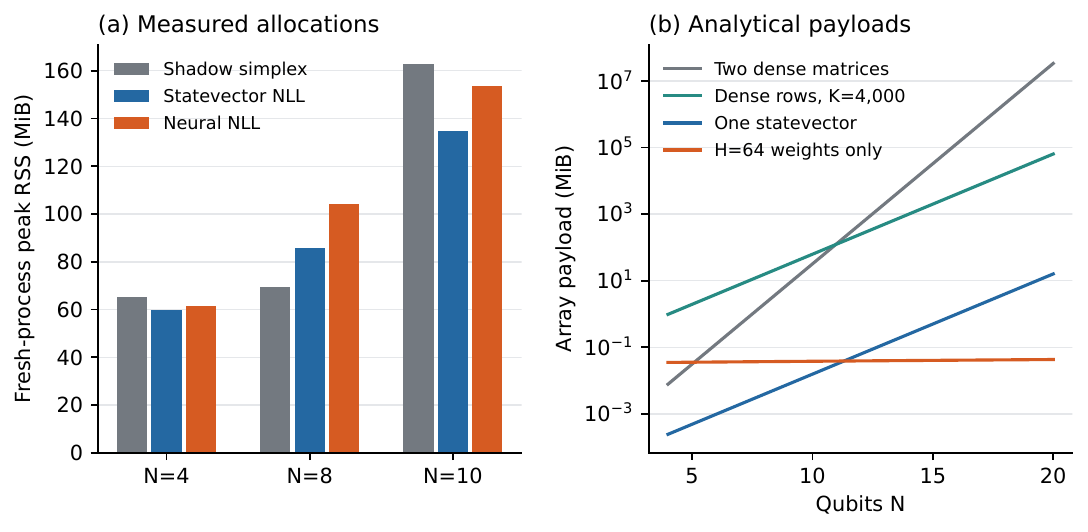}
\caption{Measured fresh-process allocation profiles and separate analytical
array-size calculations. Profiled likelihood fits use 12 L-BFGS iterations
and one restart. The right panel extrapolates formulas and does not represent
executed tomography at larger sizes.}\label{fig:resources}
\end{figure}

Figure~\ref{fig:resources} also shows analytical array sizes. For example,
at $N=20$ a single complex128 vector requires 16 MiB, two dense complex128
matrices require 32 TiB, and 4,000 dense measurement rows would require
62.5 GiB. A width-64 model has 5,634 real parameters, but that count omits
the vector, activations, likelihood contractions, and optimizer allocations.
The sparse representation improves the row-storage cost in the tested
implementation while retaining exponential dependence on $N$. No 20-qubit
state reconstruction was performed.

\section{Measurement-Aligned Phase Parameterization}
\label{sec:phaseaware}
The phase-aware branch targets the canonical GHZ coherence through a
measurement-derived phase estimate. Its architecture and fixed screen were
developed in earlier pilots, whose complete records are archived. The
main evaluation is the independent, uniform-budget 30-case cohort in
Table~\ref{tab:ghz_results}, with the same optimizer-step limits for the
generic and phase-aware neural methods.

The phase estimate used by the phase-aware branch is not supplied by the
simulator's target vector. For an all-$X/Y$ shot $i$, let $s_i$ be the
measured parity and $k_i$ the number of $Y$ axes. Under the canonical GHZ
model, its phase-dependent likelihood factor is
\begin{equation}
 \Pr(s_i|\mathbf b_i,\varphi)
 =\frac12\left[
 1+s_i\cos\left(\varphi-\frac{k_i\pi}{2}\right)
 \right].
 \label{eq:phase-aware-parity}
\end{equation}
The data-only anchor is the maximizer of the product of these factors over
the all-$X/Y$ subset. It is therefore a one-dimensional maximum-likelihood
estimate under the equal-amplitude GHZ hypothesis. It is not an estimate
of the phase of an arbitrary state; it is invoked only after the record
passes the GHZ-compatibility screen. The derivation and Fisher-information
interpretation are given in Section~\ref{sec:problem}.

\subsection{Data-only GHZ preselection and phase estimation}
For each training record, let $b_{ij}\in\{X,Y,Z\}$ denote the basis used on
qubit $j$ in shot $i$, and let $z_{ij}\in\{0,1\}$ denote the observed outcome.
We compute the fraction $S_Z$ of shots containing at least two $Z$ measurements
for which all observed $Z$ outcomes agree. We also count the shots measured
entirely in $X/Y$. The pilot classifier declares a record GHZ-like when
\begin{equation}
 S_Z\geq 0.85,\qquad M_{XY}\geq 20.
\end{equation}
For accepted records, a relative phase $\widehat\varphi$ is estimated by
maximizing the restricted equal-amplitude GHZ likelihood over all four phase
quadrants using only all-$X/Y$ parity observations. The target state and
held-out data are not inputs to either preselection or phase estimation.

On the historical pilot calibration records, the rule accepted 30/30 GHZ records
and 0/25 Haar records. These counts are a finite, idealized calibration check,
not a general classifier accuracy estimate. In particular, they do not
establish performance under readout error, depolarizing noise, local-unitary
rotations of the GHZ state, or other correlated state families. The proposed fallback for a rejected record is the generic model; the
packaged phase-aware runner rejects the record and instructs the caller to
use that alternative. The complete adaptive pipeline has not been benchmarked
on an independent mixture of state families. In particular, $S_Z$ alone is
not a coherence or entanglement certificate: an incoherent equal mixture of
$|0^N\rangle$ and $|1^N\rangle$ also has $S_Z=1$. The present branch relies
on the ideal pure-state modeling assumption as well as its parity data.

\subsection{Bounded phase-anchored polar architecture}
The original polar network predicts log-amplitude and phase using a shared
hidden trunk. The follow-up retains that parameterization and parameter count,
but replaces the unconstrained phase output by a data-derived endpoint anchor
and a bounded trainable residual:
\begin{equation}
 u_\theta(x)=\exp\!\left[
 s_\theta(x)+i\left\{
 \widehat\varphi\,\mathbf 1[x=1^N]
 +\delta_{\max}\tanh(r_\theta(x))
 \right\}\right],
 \qquad \delta_{\max}=0.30.
 \label{eq:phaseanchored}
\end{equation}
The physical wavefunction is $\psi_\theta=u_\theta/\|u_\theta\|_2$.
Here $s_\theta$ is the clipped log-amplitude output and $r_\theta$ is the
unbounded pre-activation used for the phase residual. The correction at each
basis string lies in $(-\delta_{\max},\delta_{\max})$; the relative correction
between the two GHZ endpoints can therefore span up to $2\delta_{\max}=0.60$
radians, not only 0.30 radians. The anchor $\widehat\varphi$ is fixed during
optimization, while the residual can refine the phase at every computational
basis string within the stated bound. For $N=10$ and width 128, the model
retains 18,178 trainable parameters, equal to the original polar MLP.
The phase-output weights are initialized with a small nonzero scale of 0.02;
exactly zero initialization was less robust in preliminary checks because it
suppressed phase-gradient flow through the shared trunk at initialization.

This is a structured inductive bias for canonical GHZ states in the
computational basis. It fixes an estimated phase and restricts the remaining
phase freedom; equal trainable parameter counts do not make its hypothesis
class or prior information identical to those of the generic network. It is not a general-purpose architecture for arbitrary
entangled states. Its validity depends on the data-only preselector and the
adequacy of the canonical GHZ phase model.

\subsection{Evaluation on the uniform-budget cohort}
At 8,000 shots, the mean phase-aware fidelities at four, eight, and ten
qubits are 0.9995, 0.9972, and 0.9972, respectively
(Table~\ref{tab:ghz_results}). The selected phase-aware reconstruction has
higher fidelity than the selected generic neural reconstruction in every
one of the 30 paired cases. The different phase initialization and bounded
phase family are components of the combined method; equal widths and
optimizer-step limits do not isolate their individual effects.

The GHZ-compatibility screen accepts every record in this independent
cohort. The earlier calibration check accepted 30 GHZ records and rejected
25 Haar records. Its figure and per-record scores are included in the
archive; this finite ideal-data check is not a general classifier guarantee.
The direct GHZ phase-only control remains in
Appendix~\ref{app:uniform8000-diagnostics}, so accuracy due to the exact
support and phase assumption is not attributed solely to neural training.

\subsection{Scope and extensions}
The model is evaluated on ideal canonical GHZ states. The data-only screen
and bounded phase residual assume that computational-basis structure; they
are not an entanglement witness or a general mixed-state model. Generic and
phase-aware neural optimizer limits are matched in the uniform-budget
cohort, while the strengthened vector baseline has more starts. Distinguishing
the effects of phase initialization and phase restriction requires an
additional architecture ablation. Noise, locally rotated cat states,
unequal populations, and additional coherent amplitudes are natural next
tests for the flexible model. The probability of a phase-sensitive setting
still decreases as $(2/3)^N$ under uniform local Pauli sampling.

\section{Discussion}
The results support a measurement-alignment interpretation of neural
tomography. The estimator is not defined by its network architecture alone:
it is the combination of a measurement model, a state-family prior, a
parameterization, a likelihood objective, and an optimization/selection
procedure. The experiments show why the gains due to a pure-state prior,
likelihood fitting, spectral postprocessing, neural parameterization,
measurement budget, and optimization protocol must be separated.
A normalized unrestricted vector already has the same Hermiticity, positivity,
trace, and rank-one guarantees as the neural state. Both avoid explicit
inverse-channel arithmetic during fitting. A neural advantage must therefore
be demonstrated against that direct comparator, with a clearly stated target
class and optimization budget.

The present neural models are not compressed representations of the demonstrated
Haar states. At ten qubits the Cartesian model has 69,122 real parameters,
whereas an unrestricted pure state has 2,046 real degrees of freedom. At four
qubits even the width-64 model has 4,610 parameters for 30 degrees of freedom.
Overparameterization may affect optimization or generalization, but it does not
establish wavefunction compression. Conversely, a smaller fixed-width model
could compress a restricted family without covering generic Haar states.

Best-achieved fits are meaningful only with a declared objective and
candidate budget. We use training NLL for all fitted methods, keep every
target in the aggregate, and report initialization counts rather than
selecting by simulated fidelity. Both vector and neural optimization can
benefit from alternative starts; the archive retains their full diagnostics.
For Haar states, the added ten-qubit tests do not reproduce the dramatic
recovery seen for isolated GHZ phase errors. More iterations reuse the same
observations and must be distinguished from collecting additional copies.

The uniform-budget cohort compares all methods on identical records within
a case, rather than selecting examples with favorable target fidelity.
At ten qubits the phase-aware mean is 0.9972 versus 0.9845 for the generic
network. The direct equal-amplitude GHZ phase control reaches 0.9980 on
those same records (Table~\ref{tab:uniform8000-control}), showing that much
of the accuracy on exact GHZ states is already available from their
restricted support and measured phase. The neural method is therefore a
flexible phase-regularized fit, not a necessary replacement for a
one-parameter estimator on this exact family. Its added value beyond that
control should be assessed on departures from the ideal GHZ model.

The experiments have several explicit limits. Five target/data replicates
per original condition and ten per new uniform-budget condition remain
small empirical samples; the denoiser study has only
one training seed. No mixed states, measurement miscalibration, readout noise,
or experimental-device drift are modeled. The GHZ family is structured and
does not represent generic entanglement complexity. Architecture depth, width,
and optimizer schedules are fixed rather than exhaustively tuned. Finite-budget vector and neural
fits may remain optimization-limited. These facts preclude a universal ranking
of neural architectures or a general claim that non-neural methods dominate.

Runtime and memory likewise depend on implementation. This reference uses
explicit vectors, sparse measurement rows, analytic NumPy gradients, and dense
matrices for the shadow baselines. It does not test tensor-network contractions,
GPU training, specialized likelihood solvers, or the cited literature's full
range of measurement ensembles. Polynomial-size parameter vectors do not by
themselves yield polynomial-time cross-basis likelihood evaluation. Extending
the present result to large systems requires both an appropriate restricted
state family and an efficient probability-evaluation method.

\section{Conclusions}
We introduced a measurement-aligned neural tomography strategy that
incorporates phase-sensitive measurement information into a bounded
residual phase parameterization. The completed uniform-budget comparison
uses 30 independent GHZ target/data records, 8,000 shared measurements per
case, matched generic/phase-aware neural step limits, and a strengthened
multistart statevector reference.

At ten qubits, the phase-aware model reaches mean squared fidelity
$0.9972\pm0.0030$, compared with $0.9845\pm0.0084$ for the generic neural
model and $0.9597\pm0.0039$ for the unrestricted vector. Phase-aware fitting
improves on generic neural fitting in all 30 paired cases across the three
sizes. These are training-likelihood-selected reconstructions; no target
state is used to choose a candidate. The direct GHZ phase estimator also
performs strongly, clarifying the value of the measured phase and known
support independently of the neural representation.

The complementary Haar comparison reaches mean fidelity 0.9649 at ten
qubits and 32,000 copies for both likelihood models. Additional starts and
tighter continuation at 8,000 copies yield only modest changes, separating
this sample-budget regime from isolated initialization-sensitive GHZ
outcomes. The study therefore reports achieved accuracy under explicit
measurement and optimization budgets, with the same selection principle
for both representations.

The broader contribution is a concrete design strategy: identify the
measurement statistics carrying a defining quantum feature, encode that
information into a flexible state representation, and compare the result
with both generic and structured estimators. The high-fidelity GHZ
reconstructions establish this strategy in the tested ideal setting and
motivate its extension to richer structured states and realistic
measurement conditions.

\section*{Author Contributions}
Conceptualization, X.W.;
methodology and workflow design, X.W.;
software, J.N., Y.C. and Y.W.;
formal analysis and mathematical derivations, J.N., Y.C. and Y.W.;
investigation and numerical simulations, J.N., Y.C. and Y.W.;
validation and results analysis, X.S. and Y.W.;
writing---original draft preparation, J.N., Y.C. and Y.W.;
writing---review and editing, X.S. and Y.W.;
supervision, X.W.
All authors have read and agreed to the published version of the manuscript.

\section*{Funding}
This work was supported by the State Key Laboratory of Processors, Institute of Computing Technology, Chinese Academy of Sciences, under Grant No. CLQ202404, the Beijing Natural Science Foundation (Grant No. 4262015) and the São Paulo Research Foundation (FAPESP), Brazil (Process Nos. 2025/25586-5 and 2021/02120-0).

\section*{Institutional Review Board Statement}
Not applicable. The study consists entirely of numerical simulations and does not involve human participants or animals.

\section*{Informed Consent Statement}
Not applicable.

\section*{Data Availability Statement}
The reproducibility archive contains the original primary
records, the 30-case uniform GHZ cohort, original and additional fitted
states, candidate summaries, protocols, source code, and numerical audits.
Historical tables and initialization diagnostics are retained separately
from the main manuscript. Execution records document the actual processor,
software, and concurrency settings for each cohort.


\section*{Conflicts of Interest}
The authors declare no conflicts of interest.

\section*{Abbreviations}
The following abbreviations are used in this manuscript:\\
\noindent
\begin{tabular}{@{}ll}
GHZ & Greenberger--Horne--Zeilinger\\
MLE & Maximum-likelihood estimation\\
NLL & Negative log likelihood\\
POVM & Positive-operator-valued measure\\
QST & Quantum-state tomography
\end{tabular}

\appendix
\section{Structured control and reproducibility}
\label{app:uniform8000-diagnostics}
The direct phase estimator fixes the exact GHZ support and equal populations
and uses the same training-derived phase as the neural anchor. It is a
structured control, not an unrestricted wavefunction estimator.


\begin{table}[H]
\small
\setlength{\tabcolsep}{4pt}

\caption{Structured control on the same ten records per size.
The direct estimator fixes the exact GHZ support and equal endpoint
populations, using only the training-derived phase estimate.
The final column is the paired phase-aware-minus-direct fidelity
difference (mean $\pm$ sample SD). These exact-GHZ data do not require
a neural model to achieve high fidelity.}
\label{tab:uniform8000-control}

\centering
\begin{tabular}{@{}rccc@{}}
\toprule
$N$
& Direct GHZ phase estimate
& Phase-aware neural
& Paired $\Delta F$ \\
\midrule
4
& $0.9999\pm 0.0001$
& $0.9995\pm 0.0003$
& $-0.00035\pm 0.00024$ \\
8
& $0.9975\pm 0.0040$
& $0.9972\pm 0.0034$
& $-0.00032\pm 0.00105$ \\
10
& $0.9980\pm 0.0031$
& $0.9972\pm 0.0030$
& $-0.00087\pm 0.00051$ \\
\bottomrule
\end{tabular}

\end{table}

The archive retains available histories and candidate summaries for both
fitted methods. It also contains the historical lower-budget tables and
diagnostic source sections. Every independent target/data record enters
the main tables, with one fit selected per record by minimum training NLL.

The Haar audit checks saved parameters against wavefunctions, recomputes
losses and fidelities, and verifies data hashes and restart selection.
Independent numerical tests check gradients and measurement rows. The
protocol, fixed seeds, forty additional fits, and numerical summaries are
retained together with the uniform-budget GHZ verification records.
Numerical comparisons are not global-optimality certificates; execution
settings and timing hardware are recorded separately from the algorithm.

\clearpage

\end{document}